\documentclass{JINST}

\usepackage{longtable}
\usepackage{multirow} %para crear tablas donde el número de filas no es constante
\usepackage{lscape}
\usepackage{lineno}
\title{Radiopurity control in the NEXT-100 double beta decay experiment: procedures and initial measurements}
\author{
V.~\'Alvarez,$^{a}$ I.~Bandac,$^{b}$ A.~Bettini,$^{b,c}$
F.I.G.M.~Borges,$^{d}$ S.~C\'arcel,$^{a}$ J.~Castel,$^{b,e}$
S.~Cebri\'an,$^{b,e}$\thanks{Corresponding author
(scebrian@unizar.es).}~ A.~Cervera,$^{a}$ C.A.N.~Conde,$^{d}$
T.~Dafni,$^{b,e}$ T.H.V.T.~Dias,$^{d}$ J.~D\'iaz,$^{a}$
M.~Egorov,$^{f}$ R.~Esteve,$^{g}$ P.~Evtoukhovitch,$^{h}$
L.M.P.~Fernandes,$^{d}$ P.~Ferrario,$^{a}$ A.L.~Ferreira,$^{i}$
E.D.C.~Freitas,$^{d}$ V.M.~Gehman,$^{f}$ A.~Gil,$^{a}$
A.~Goldschmidt,$^{f}$ H.~G\'omez,$^{b,e}$\thanks{Present address:
Laboratoire de l'Accélérateur Linéaire (LAL). Centre Scientifique
d'Orsay. Bâtiment 200 - BP 34. 91898 Orsay Cedex, France.}~
J.J.~G\'omez-Cadenas,$^{a}$\thanks{Spokesperson
(gomez@mail.cern.ch).}~ D. Gonz\'alez-D\'iaz,$^{b,e}$
R.M.~Guti\'errez,$^{j}$ J.~Hauptman,$^{k}$ J.A.~Hernando
Morata,$^{l}$ D.C.~Herrera,$^{b,e}$ F.J.~Iguaz,$^{b,e}$
I.G.~Irastorza,$^{b,e}$ M.A.~Jinete,$^{j}$ L.~Labarga,$^{m}$
A.~Laing,$^{a}$ I.~Liubarsky,$^{a}$ J.A.M.~Lopes,$^{d}$
D.~Lorca,$^{a}$ M.~Losada,$^{j}$ G.~Luz\'on,$^{b,e}$
A.~Mar\'i,$^{g}$ J.~Mart\'in-Albo,$^{a}$ A.~Mart\'inez,$^{a}$
T.~Miller,$^{f}$ A.~Moiseenko,$^{h}$ F.~Monrabal,$^{a}$
C.M.B.~Monteiro,$^{d}$ F.J.~Mora,$^{g}$ L.M. Moutinho,$^{i}$
J.~Mu\~noz Vidal,$^{a}$ H.~Natal da Luz,$^{d}$ G.~Navarro,$^{j}$
M.~Nebot-Guinot,$^{a}$ D.~Nygren,$^{f}$ C.A.B.~Oliveira,$^{f}$ A.
Ortiz de Sol\'orzano,$^{b,e}$ R.~Palma,$^{n}$ J.~P\'erez,$^{o}$
J.L.~P\'erez Aparicio,$^{n}$ J.~Renner,$^{f}$ L.~Ripoll,$^{p}$
A.~Rodr\'iguez,$^{b,e}$ J.~Rodr\'iguez,$^{a}$ F.P.~Santos,$^{d}$
J.M.F.~dos Santos,$^{d}$ L.~Segui,$^{b,e}$ L.~Serra,$^{a}$
D.~Shuman,$^{f}$ A. Sim\'on,$^{a}$ C.~Sofka,$^{q}$ M.~Sorel,$^{a}$
J.F.~Toledo,$^{g}$ A.~Tom\'as,$^{b,e}$ J.~Torrent,$^{p}$
Z.~Tsamalaidze,$^{h}$ D.~V\'azquez,$^{l}$ J.F.C.A.~Veloso,$^{i}$
J.A.~Villar,$^{b,e}$ R.C.~Webb,$^{q}$ J.T.~White$^{q}$ and
N.~Yahlali$^{a}$

\\
\llap{$^{a}$}
Instituto de F\'isica Corpuscular (IFIC), CSIC \& Universitat de Val\`encia\\
Calle Catedr\'atico Jos\'e Beltr\'an, 2, 46980 Paterna, Valencia, Spain\\
\llap{$^b$}
Laboratorio Subterráneo de Canfranc\\
Paseo de los Ayerbe s/n, 22880 Canfranc Estación, Huesca, Spain\\
\llap{$^{c}$}
Padua University and INFN Section, Dipartimento di
Fisca G. Galilei, Via Marzolo 8, 35131 Padova, Italy\\
\llap{$^{d}$}
Departamento de Fisica, Universidade de Coimbra\\
Rua Larga, 3004-516 Coimbra, Portugal\\
\llap{$^{e}$}
Laboratorio de F\'isica Nuclear y Astropart\'iculas, Universidad de Zaragoza\\
Calle Pedro Cerbuna 12, 50009 Zaragoza, Spain\\
\llap{$^{f}$}
Lawrence Berkeley National Laboratory (LBNL)\\
1 Cyclotron Road, Berkeley, California 94720, USA\\
\llap{$^{g}$}
Instituto de Instrumentaci\'on para Imagen Molecular (I3M), Universitat Polit\`ecnica de Val\`encia\\
Camino de Vera, s/n, Edificio 8B, 46022 Valencia, Spain\\
\llap{$^{h}$}
Joint Institute for Nuclear Research (JINR)\\
Joliot-Curie 6, 141980 Dubna, Russia\\
\llap{$^{i}$}Institute of Nanostructures, Nanomodelling and Nanofabrication (i3N), Universidade de Aveiro\\
Campus de Santiago, 3810-193 Aveiro, Portugal\\
\llap{$^{j}$}
Centro de Investigaciones en Ciencias B\'asicas y Aplicadas, Universidad Antonio Nari\~no\\
Carretera 3 este No.\ 47A-15, Bogot\'a, Colombia\\
\llap{$^{k}$}
Department of Physics and Astronomy, Iowa State University\\
12 Physics Hall, Ames, Iowa 50011-3160, USA\\
\llap{$^{l}$}
Instituto Gallego de F\'isica de Altas Energ\'ias (IGFAE), Univ.\ de Santiago de Compostela\\
Campus sur, R\'ua Xos\'e Mar\'ia Su\'arez N\'u\~nez, s/n, 15782 Santiago de Compostela, Spain\\
\llap{$^{m}$}
Departamento de F\'isica Te\'orica, Universidad Aut\'onoma de Madrid\\
Campus de Cantoblanco, 28049 Madrid, Spain\\
\llap{$^{n}$}
Dpto.\ de Mec\'anica de Medios Continuos y Teor\'ia de Estructuras, Univ.\ Polit\`ecnica de Val\`encia\\
Camino de Vera, s/n, 46071 Valencia, Spain\\
\llap{$^{o}$}
Instituto de F\'isica Te\'orica (IFT), UAM/CSIC\\
Campus de Cantoblanco, 28049 Madrid, Spain\\
\llap{$^{p}$}
Escola Polit\`ecnica Superior, Universitat de Girona\\
Av.~Montilivi, s/n, 17071 Girona, Spain\\
\llap{$^{q}$}
Department of Physics and Astronomy, Texas A\&M University\\
College Station, Texas 77843-4242, USA\\

}

\abstract{The ``Neutrino Experiment with a Xenon Time-Projection
Chamber'' (NEXT) is intended to investigate the neutrinoless double
beta decay of $^{136}$Xe, which requires a severe suppression of
potential backgrounds. An extensive screening and material selection
process is underway for NEXT since the control of the radiopurity
levels of the materials to be used in the experimental set-up is a
must for rare event searches. First measurements based on Glow
Discharge Mass Spectrometry and gamma-ray spectroscopy using
ultra-low background germanium detectors at the Laboratorio
Subterráneo de Canfranc (Spain) are described here. Activity results
for natural radioactive chains and other common radionuclides are
summarized, being the values obtained for some materials like copper
and stainless steel very competitive. The implications of these
results for the NEXT experiment are also discussed.}

\begin{document}

%\maketitle
\section{Introduction}

\begin{figure}
\begin{center}
  \includegraphics[height=.4\textheight]{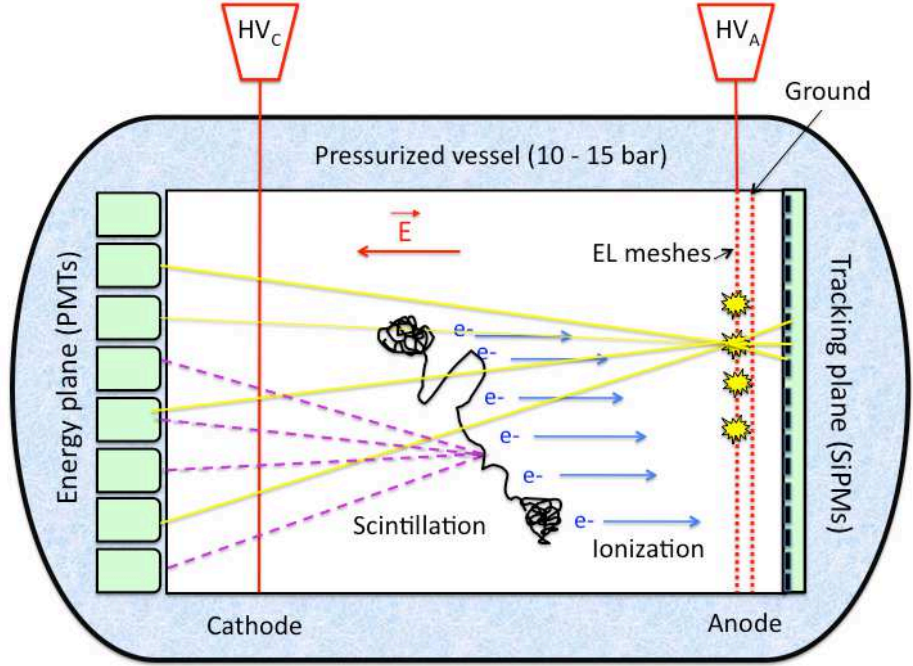}
  \caption{Concept of the NEXT experiment: light from the Xe electroluminescence (EL) generated
at the anode is recorded both in the photosensor plane right behind
it for tracking and in the photosensor plane behind the transparent
cathode for a precise energy measurement. Primary scintillation
defining the start of the event is also detected by the cathode
photosensors.}
  \label{soft}
\end{center}
\end{figure}

%%%%%% DBD and NEXT
The observation of neutrinoless double beta decay as a peak in the
sum energy of the two emitted electrons ($Q_{\beta\beta}$), would
show that neutrinos are Majorana particles and contribute to the
determination of their mass hierarchy (see for instance
\cite{dbdrefs}). The NEXT experiment (``\underline{N}eutrino
\underline{E}xperiment with a \underline{X}enon
\underline{T}ime-Projection Chamber'') will search for such a decay
in $^{136}$Xe using a high-pressure gaseous xenon Time-Projection
Chamber (TPC) with a source mass of the order of 100 kg at the
Laboratorio Subterráneo de Canfranc (LSC) \cite{lsc}, located at the
Spanish Pyrenees. The NEXT-100 detector design \cite{jinsttdr} is
intended to combine, keeping the detector$=$source approach, the
measurement of the topological signature of the event (to
discriminate the signal from background) with the energy resolution
optimization; this is possible thanks to the use of proportional
electroluminescent (EL) amplification. As illustrated in figure
\ref{soft}, energy and tracking readout planes are located at
opposite sides of the pressure vessel using different sensors:
photomultiplier tubes (PMTs) for calorimetry (and for fixing the
start of the event) and silicon multi-pixel photon counters (MPPCs)
for tracking. Energy resolution below 1\% FWHM at the
$Q_{\beta\beta}$ energy seems reachable \cite{jinsttdr}. The
expected sensitivity of NEXT is very competitive; electron neutrino
effective Majorana masses below 100 meV could be explored for a
total exposure of 500 kg$\cdot$year \cite{sense}. The Technical
Design Report of the experiment was presented in 2011, while work on
prototypes is ongoing \cite{protos}.

%%%%%% Background reduction
Operating in deep underground locations and using both active and
passive shields is mandatory in experiments searching for rare
phenomena like the nuclear double beta decay, the direct detection
of dark matter and the interaction of low-energy neutrinos
\cite{heusser,formaggio}. A careful selection of radiopure materials
to be used in the experimental set-up based on precise measurements
of ultra-low radioactivity is also compulsory in these experiments
in order to achieve, for example in NEXT, a background level of
$8\times10^{-4}$ counts keV$^{-1}$ kg$^{-1}$ y$^{-1}$ in the energy
region of interest (RoI) \cite{jinsttdr}.
%%%%% Previous results on material screening
A great effort has been done by several collaborations to measure
the radiopurity of many different materials (see for instance Refs.
\cite{ros99}-\cite{kimballton}) and information has been compiled
and made public \cite{UKDMC,ILI11}. Although these data are
extremely useful and can be used as a starting point in the
selection of materials for any experiment, a thorough radiopurity
control of the relevant components to be actually used must be
always undertaken since radiopurity requirements are stringent.

%%%%% NEXT background
The ability to discriminate signal from background is a powerful
tool in NEXT. Signal events will appear uniformly distributed in the
source volume of enriched xenon and will have a distinctive topology
(a twisted long track, about 30 cm long at 10 bar, with blobs at
both ends). Defining a fiducial volume eliminates all charged
backgrounds entering the detector while confined tracks generated by
neutral particles, like high-energy gammas, can be suppressed by
pattern recognition. Thus, the relevance of a background source
depends on its probability of generating a signal-like track in the
active volume with energy around Q$_{\beta\beta}$. The neutrinoless
double beta decay of $^{136}$Xe would produce a peak at
$Q_{\beta\beta}=2.458$\,MeV \cite{redshaw}; the main background
sources at the NEXT experiment are the gamma lines at 2.615\,MeV
from $^{208}$Tl and at 2.448\,MeV from $^{214}$Bi. These isotopes
produced at the lower part of the natural radioactive chains of
$^{232}$Th and $^{238}$U respectively.
\begin{itemize}
\item For the 2.615\,MeV line, the Compton edge is well below
Q$_{\beta\beta}$, but a scattered gamma can interact and produce
other electron tracks close enough to the initial Compton electron,
so they are reconstructed as a single object falling in the RoI.
These photons can also be scattered outside the detector and then
suffer photoelectric absorption inside contributing to the RoI. In
addition, photoelectric electrons from the 2.615\,MeV emission are
produced above the RoI but can lose energy via bremsstrahlung and
populate the energy window, if the emitted photons escape out of the
detector. \item The gamma line at 2.448 \,MeV is dangerous in spite
of its low intensity (1.57\%) because it is very close to
Q$_{\beta\beta}$.
%After the decay of
%$^{214}$Bi, other some gamma lines above this energy are emitted too
%and their Compton spectra could produce background tracks in the
%ROI, but they have lower intensity (below 0.1\%).
\end{itemize}

Both $^{208}$Tl and $^{214}$Bi are in the progeny of Rn isotopes
($^{220}$Rn and $^{222}$Rn respectively) which are present in air,
so radon diffusion from the laboratory air as well as radon
emanation from materials must be controlled. Another possible
external source of background are neutrons, whether produced by
natural radioactivity in the walls or shielding or as secondary
products of cosmic muons; preliminary estimations seem to point that
these contributions are very much below the level of concern for
NEXT. Also high energy gammas can be produced in muon-induced
electromagnetic cascades; although they seem to be of no importance
for NEXT, they could be partially tagged by an active muon veto in
the shielding.

%%%%%% NEXT requirements
Consequently, the activity of the lower part of $^{232}$Th and
$^{238}$U chains ($^{208}$Tl and $^{214}$Bi isotopes) from the
components of the set-up and at the laboratory is the main concern
in NEXT. According to the NEXT-100 design, the list of the main
materials subject to radiopurity control and the target radiopurity
for each of them were established (see table 3 in \cite{jinsttdr}).
The relevant materials for the different parts of the experimental
set-up to be considered include: lead and copper for shielding;
stainless steel and inconel for vessel; PEEK (PolyEther Ether
Ketone), polyethylene and coating materials for High Voltage (HV)
and electroluminescence components; photomultipliers, windows,
printed circuit boards and other electronic components (capacitors,
resistors, cables\dots) for the energy and tracking planes. The
target sensitivity was fixed following results from the literature
and from the first measurements made for NEXT.

%%%%%%%%%% Tomada de 1202.0721v2
%\begin{table}
%\caption{Required activity (in ${\rm mBq}/{\rm kg}$) of the most
%relevant materials used in NEXT (taken from Table 3 at
%\cite{jinsttdr}). Activities for the lower part of the uranium
%chain, starting at $^{226}$Ra, have been quoted when possible.}
%\label{tab:RA}
%\begin{center}
%\begin{tabular}{llcll}
%\toprule Material & Subsystem & Method/Ref. &$^{238}$U & $^{232}$Th
%\\ \midrule
%%
%Lead, from Cometa & Shielding & GDMS & 0.37 & 0.07 \\
%%
%Copper, from Luvata & ICS & GDMS & $<0.012$ & $<0.004$ \\
%%
%Steel (316Ti) & PV & \cite{APR11} & $<1.9$ & $<1$ \\
%%
%Bolts Inconel 718 & PV & Ge LSC & $<5.6$ & $<4.6$ \\
%%
%Bolts Inconel 625 & PV & Ge LSC & $<1.8$ & $<2.0$ \\
%%
%PEEK, from Sanmetal & FC/EP/TP & Ge Unizar & 36.3 & 11.7 \\
%%
%Capacitors (Tantalum) & FC/EP/TP & \cite{ILI11} & 320 & 1230 \\
%
%SMD Resistors, Finechem (per pc) & FC & Ge Unizar & 0.022 & $<$0.048 \\
%%
%Polyethylene & FC & \cite{APR11} & 0.23 & $<$0.14 \\
%%
%TTX & FC & \cite{BOC09}& 12.4 & $<$1.6 \\
%%
%TPB & FC/EP/TP & \cite{SNO} & 1.63 & 0.47 \\
%%
%PTFE (Teflon) & EP/TP/DB & \cite{BUD09} & 0.025 & 0.031 \\
%%
%PMT (R11410-MOD per pc) & EP & \cite{APR11} & $< 2.5$ & $< 2.5$ \\
%%
%PMT (R11410-MOD per pc) & EP & \cite{FAH11}& $< 0.4$ & $< 0.3$ \\
%%
%Sapphire window & EP & \cite{LEO08}& $<$0.31 & 0.12 \\
%%
%CUFLON & TP & \cite{NIS09}& 0.36 & 0.28 \\
%%
%Kapton cable & TP/EP & \cite{APR11} & $<$11 & $<$11 \\
%\bottomrule
%\end{tabular}
%\end{center}
%\end{table}
%%%%%%%%%%

%%%%% Paper structure
In this paper, the initial material radiopurity control performed
for the NEXT experiment is presented. First, in Sec. \ref{tec} the
techniques and equipment used to carry out the measurements of
activity levels are described. Then, the results obtained for all
the analyzed materials are shown and discussed in Sec. \ref{res}.

\section{Techniques and equipment} \label{tec}

The techniques employed for the radiopurity measurements for NEXT
are Glow-Discharge Mass Spectrometry (GDMS) and gamma-ray
spectroscopy using ultra-low background germanium detectors operated
deep underground. Each one has advantages and drawbacks, which make
them more or less suitable depending on the context.

\subsection{GDMS}
%% technique
In GDMS, the sample to be analyzed forms the cathode in a plasma or
discharge gas, typically argon. Argon ions are accelerated toward
the sample resulting in erosion and atomization of its surface. The
sputtered species are transported into the plasma where they are
ionized. Ions are then extracted for mass spectrometry. This method
is consequently very suitable for metals. Similarly to other
techniques also based on mass spectrometry, it is fast and requires
only a small sample of the material (typically, samples with a
surface of 2$\times$2 cm$^{2}$ were prepared for NEXT measurements).
However, the output given is normally only the concentration of
elements while particular isotopes are not identified. Measured
concentrations of U, Th and K have been converted to $^{232}$Th,
$^{238}$U and $^{40}$K activities\footnote{Conversion factors are: 1
ppb U $=$ 12.4 mBq/kg $^{238}$U, 1 ppb Th = 4.1 mBq/kg $^{232}$Th
and 1 ppm $^{nat}$K = 31 mBq/kg $^{40}$K.}. Having no information on
daughter nuclides in the chains, a possible disequilibrium cannot be
detected. In any case, this technique can be very helpful to make a
pre-selection of samples to be later screened with germanium
detectors underground. GDMS measurements for NEXT materials were
made by Shiva Technologies (Evans Analytical
Group\footnote{http://www.eaglabs.com}) in France.

\subsection{Germanium gamma-ray spectrometry}

%technique
The most important advantage of gamma-ray spectrometry performed
with germanium detectors is that, being non-destructive, the actual
components to be used in experiments can be examined. Germanium
detectors offer very good energy resolution and low intrinsic
background. In addition, activities of isotopes dangerous for the
experiment background are directly assessed. No sample pre-treatment
is necessary, but massive samples and time-consuming measurements
(lasting up to several weeks) are needed to quantify very low
activities.

\begin{figure}
\begin{center}
  \includegraphics[height=.4\textheight]{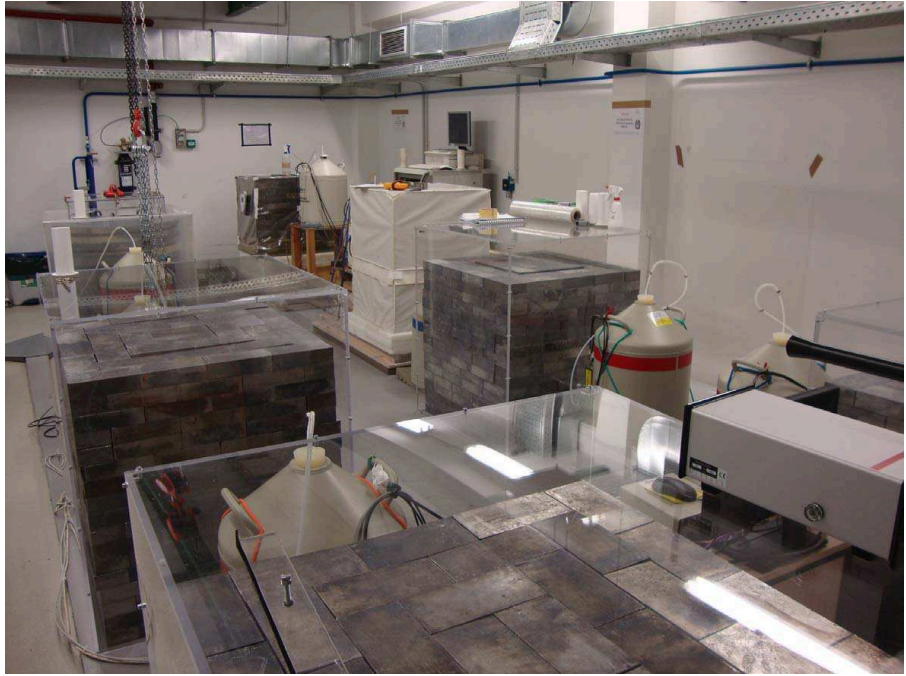}
  \caption{Picture of the Hall C of the LSC where several germanium detectors inside their shieldings are operated for material radiopurity screening (courtesy of LSC). Background counting rates are reported in table 1.}
  \label{gelsc}
\end{center}
\end{figure}

%% facilities
The LSC offers a Radiopurity
Service\footnote{http://www.lsc-canfranc.es/en/for-users/lsc-services/radiopurity.html}
to measure ultra-low level radioactivity using several germanium
detectors operated at the Hall C for gamma-ray spectroscopy (see
figure \ref{gelsc}); it is intended to support the construction of
experiments to be operated there. Since the laboratory is located at
a depth of 2450 m.w.e., the cosmic muon flux is about 5 orders of
magnitude lower than at sea-level surface. Radon activity in the air
is between 50 and 80 Bq/m$^{3}$ in the underground halls
\cite{ulabs}. All germanium spectroscopy measurements presented here
were carried out at LSC using in particular four different $\sim$2.2
kg detectors from LSC and a $\sim$1 kg detector from the University
of Zaragoza.

\begin{itemize}
\item The detectors property of LSC are p-type close-end coaxial High Purity germanium
detectors produced by Canberra France with 100-110\% relative
efficiencies\footnote{Efficiency relative to a $3''\times3''$ NaI
detector at 1332 keV and for a distance of 25 cm between source and
detector.} and a FWHM energy resolution of $\sim$2 keV at the
$^{60}$Co gamma line of 1332 keV. The active volume of crystals
ranges from 410 to 420 cm$^{3}$ and cryostats are made of ultra-low
background aluminum. The data acquisition system is based on Digital
Signal Processing using Canberra DSA1000 modules. Each detector has
a shield consisting of 5 cm of oxygen-free copper and 20 cm of very
low activity lead having $<$30 mBq/kg of $^{210}$Pb; nitrogen gas is
flushed inside a methacrylate box to avoid airborne radon intrusion.
The detectors used for NEXT measurements are those named GeOroel,
GeTobazo, GeAnayet and GeLatuca; they are in operation since 2011.
%The background counting rate from 100
%to 2700 keV in the described shielding is about 0.75 counts/minute
%for GeOroel and about 1 count/minute for GeTobazo and GeAnayet.
\item The detector from University of Zaragoza, named Paquito, is
also a p-type close-end coaxial High Purity germanium detector, but
with a smaller crystal volume of 190 cm$^{3}$ inside a copper
cryostat. The FWHM energy resolution at 1332 keV is 2.7 keV. It is
operated inside a shield made of 10 cm of archaeological lead plus
15 cm of low activity lead with nitrogen flush too. The electronic
chain for data acquisition is based on standard Canberra 2020 Linear
Amplifier and Canberra 8075 Analog-to-Digital-Converter modules.
This detector has been used for radiopurity measurements at Canfranc
for several years (more details can be found in
\cite{radiopuritymm}).
% and has a background counting rate from 100
%to 2700 keV of about 0.05 counts/minute.
\end{itemize}

\begin{figure}
\begin{center}
  \includegraphics[height=.4\textheight]{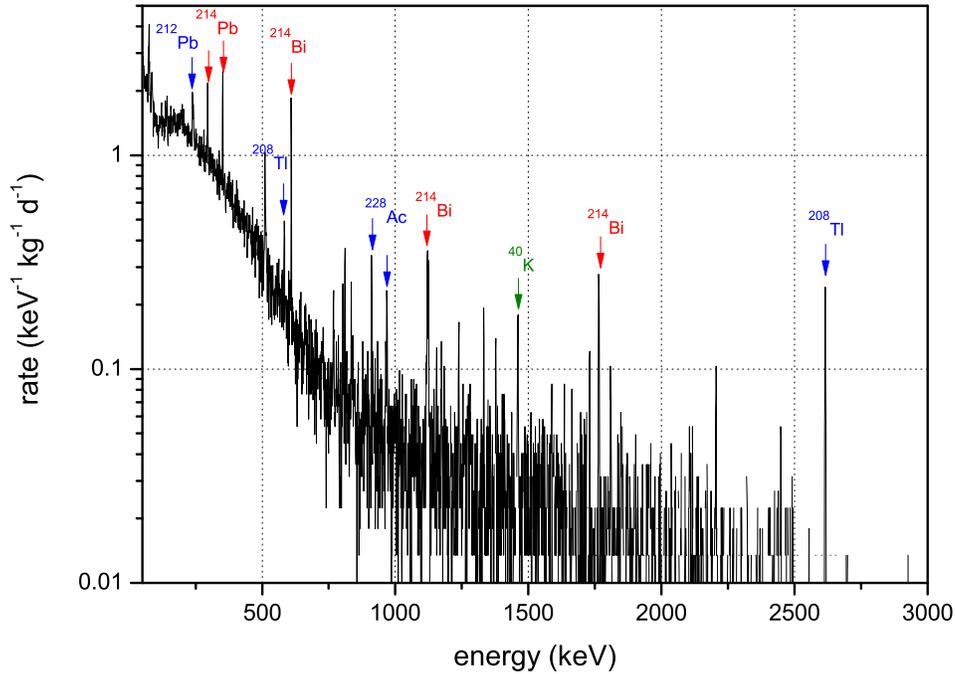}
  \caption{Energy spectrum of GeOroel detector at LSC registered in a background measurement for 38.00 days.
Main gamma lines from isotopes of the $^{238}$U (red) and $^{232}$Th
(blue) radioactive chains and from $^{40}$K (green) are marked.}
  \label{gespectrum}
\end{center}
\end{figure}

%% backgrounds
The background of each detector inside its shielding is determined
by taking data with no sample for long periods of time of at least
one month, due to the low counting rates. Table \ref{gerates} shows
the counting rates of all the detectors used in the NEXT
measurements in the energy window from 100 to 2700 keV and at
different peaks: 583 keV from $^{208}$Tl, 609 keV from $^{214}$Bi
and 1461 keV from $^{40}$K; all the rates are expressed in counts
per day and per kg of germanium detector and correspond to
background measurements performed from May to July 2012. The
complete background spectrum of GeOroel detector, registered for
38.00 days, is presented in figure \ref{gespectrum}. Lines from
cosmogenic isotopes like $^{54}$Mn, $^{58}$Co and $^{65}$Zn, having
half-lives around one year, are still observable in GeOroel detector
but not in the other ones.

\begin{table}
\caption{Background counting rates (expressed in counts d$^{-1}$
kg$^{-1}$) of the germanium detectors used at LSC for the NEXT
measurements. Integral rate from 100 to 2700 keV and rates at
different peaks (583 keV from $^{208}$Tl, 609 keV from $^{214}$Bi
and 1461 keV from $^{40}$K) are presented. Only statistical errors
are quoted.}
%\begin{center}
%\footnotesize
\begin{tabular}{lcr@{$\pm$}lccc}
\\
\hline

Detector name  & Mass (kg) &  \multicolumn{2}{c}{100-2700 keV} & 583
keV &609 keV &1461 keV  \\ \hline

GeOroel & 2.230 &   490&2 &    0.8$\pm$0.1 &    3.0$\pm$ 0.2
& 0.41$\pm$0.07\\

 GeAnayet &     2.183 &  714&3 &  3.8$\pm$0.2 &   1.7$\pm$0.1& 0.38$\pm$0.07 \\

GeTobazo  &  2.185 &  708&3  & 4.0$\pm$0.2 & 1.3$\pm$0.1 &
0.40$\pm$0.06 \\

GeLatuca  &  2.187 &  710&3  & 3.3$\pm$0.2 & 5.9$\pm$0.3 &
0.56$\pm$0.08 \\

 Paquito  & 1 &  79&2 &  0.27$\pm$0.09 & 0.5$\pm$0.1 & 0.25$\pm$0.09 \\

 \hline
\end{tabular}

\label{gerates}
%\end{center}
\end{table}

%%analysis: efficiency
To derive the activity in a sample of an isotope producing a gamma
emission of a certain energy, the main ingredients (together with
the time of the measurement and the branching ratio of the emission)
are the net signal (that is, the number of events at the gamma line
due to the sample) and the full-energy peak detection efficiency at
the corresponding energy.

In the context of determining ultra-low activities of a sample (at
the level of mBq/kg and below), when comparing the measured energy
spectrum of the sample with that of the detector background, it is
not straightforward to decide if the gross signal can be considered
to statistically differ from the background signal. The criteria
proposed in Currie's landmark paper \cite{currie} and more recently
revised in \cite{gator,revisiting} have been followed here.
Activities have been quantified when possible and upper limits with
a 95.45\% C.L. have been derived otherwise.

\begin{figure}
\begin{center}
  \includegraphics[height=.4\textheight]{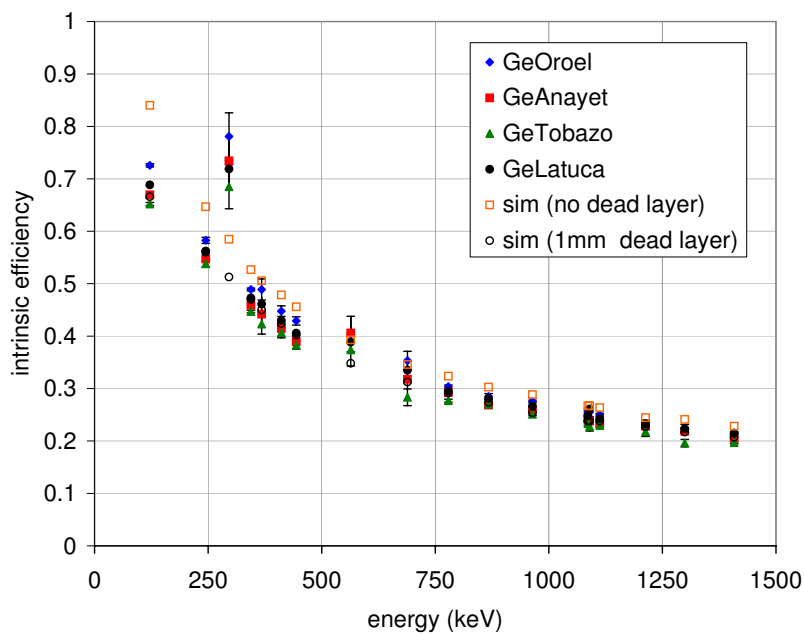}
  \caption{Intrinsic efficiency measured for different Ge detectors using a $^{152}$Eu reference source and the corresponding simulation, considering a 1-mm-thick crystal dead layer or no dead layer.}
  \label{efcurves}
\end{center}
\end{figure}

Concerning the estimate of the detection efficiency, Monte Carlo
simulations based on the Geant4 \cite{geant4} code have been
performed for each sample, accounting for intrinsic efficiency, the
geometric factor and self-absorption at the sample. No relevant
change has been observed in the Geant4 simulation when changing
version or the physical models implemented for interactions
(considering the low energy extensions for electromagnetic processes
based on theoretical models and on exploitation of evaluated data,
G4EmLivermorePhysics class and the previous G4LowEnergy* classes
\cite{geant4physics}). Validation of the simulation has been made by
comparing the efficiency curve of the detectors measured with a
$^{152}$Eu reference source of known activity located at 25 cm from
the detector with the simulated one. Figure \ref{efcurves} shows the
intrinsic efficiency (corrected by solid angle) obtained for
GeOroel, GeTobazo, GeAnayet and GeLatuca detectors together with the
corresponding simulations for GeTobazo/GeAnayet. The higher
efficiency shown by GeOroel is due to a slightly larger volume in
comparison with the other detectors. The inclusion of a 1-mm-thick
dead layer in the simulation has improved the agreement with
measurements, especially at low energies, reducing deviations to a
level of 5\%; an overall uncertainty of 10\% is considered for the
simulated detection efficiency of the samples and propagated to the
final activity value. The larger discrepancy observed in figure
\ref{efcurves} for 295 keV values is due to the interference in the
measurements at this energy of the gamma line of $^{214}$Pb,
descendant of $^{222}$Rn present in air (measurements with the
$^{152}$Eu source were carried out with the shields partially open
and no background substraction was made). The relative efficiencies
derived from these measurements and also from the simulation
reproduce the values specified by the manufacturer.

Activities of different sub-series in the natural chains of
$^{238}$U, $^{232}$Th and $^{235}$U as well as of common primordial,
cosmogenic or anthropogenic radionuclides like $^{40}$K, $^{60}$Co
and $^{137}$Cs have been evaluated by analyzing the most intense
gamma lines of different isotopes. Outgassing and chemical
procedures in materials can make secular equilibrium in radioactive
chains break, so information provided by germanium detectors at the
different stages is very important. For $^{238}$U, emissions from
$^{234}$Th and $^{234m}$Pa are searched to quantify activity of the
upper part of the chain and lines from $^{214}$Pb and $^{214}$Bi for
the sub-chain starting with $^{226}$Ra up to $^{210}$Pb. For
$^{232}$Th chain, emissions of $^{228}$Ac are analyzed for the upper
part and those of $^{212}$Pb, $^{212}$Bi and $^{208}$Tl for the
lower one. Concerning $^{235}$U chain, only emissions from the
parent isotope are taken into account since all the other ones have
very low intensities.

\section{Measurements and results} \label{res}

Materials of very massive components and of those to be located
inside the detector vessel have been selected for the first
screening campaign of NEXT, trying to keep up with the schedule of
construction of several subsystems like the pressure vessel, the
field cage and the boards for the tracking plane. The performed
measurements are described and discussed in the context of NEXT.
Details concerning the detector used, the size of the sample and the
time of data taking for all measurements using germanium detectors
are presented in table \ref{gedata}. The results obtained are all
independently summarized in table \ref{rpm}; reported errors
correspond to $1\sigma$ uncertainties including both statistical and
efficiency uncertainties.

\begin{table}
\caption{Information on the germanium gamma-ray spectrometry
measurements performed for NEXT at LSC: material, detector used,
samples size (mass, area or number of pieces) and screening time.
The corresponding row number of table 3 where the activity values
obtained for each sample are reported is also quoted.}
%\begin{center}
%\footnotesize
\begin{tabular}{lccr@{ }lr@{.}l}
\\
\hline
Material, Supplier & \# in table 3 &  Detector & \multicolumn{2}{c}{Sample size} &  \multicolumn{2}{c}{Screening time (d)} \\
\hline

Pb, Tecnibusa & 5 & GeAnayet &  5585&g & 19&44 \\
Pb, Tecnibusa & 6 & GeAnayet &  5585&g& 35&99 \\
Cu, Luvata & 10 & Paquito & 681&g& 39&17  \\
Ti, SMP & 11 & GeOroel & 121&g& 38&46 \\
Ti, SMP & 12 & GeTobazo & 121&g& 43&11\\
Ti, Ti Metal Supply & 13 & GeOroel & 1804&g& 47&23\\
304L Stainless Steel, Pfeiffer & 14 & Paquito & 347&g& 19&55 \\
316Ti Stainless Steel, 10 mm, Nironit & 15 & GeTobazo & 7684&g& 33&00 \\
316Ti Stainless Steel, 15 mm, Nironit & 16 & GeTobazo & 10205&g & 35&61\\
316Ti Stainless Steel, 50 mm, Nironit & 17 & GeAnayet & 4816&g & 34&72\\
Inconel 625, Mecanizados Kanter & 18 & GeTobazo & 1004&g & 27&98 \\
Inconel 718, Mecanizados Kanter & 19 & GeOroel & 611&g& 27&93 \\
PEEK, Sanmetal & 20& Paquito & 459&g& 24&27 \\
Polyethylene, IN2 Plastics & 21& GeAnayet & 1315&g & 36&76 \\
Semitron ES225, Quadrant EPP & 22& GeOroel & 1618&g & 35&05  \\
SMD resistor, Farnell & 23 & Paquito & 50&pc & 18&15 \\
SM5D resistor, Finechem & 24 & Paquito & 100&pc & 31&45 \\
Kapton-Cu PCB, LabCircuits& 25 & Paquito &260.15&cm$^{2}$ & 35&28
\\
Cuflon, Polyflon & 26 & GeOroel & 1876&g & 24&29 \\
Bonding films, Polyflon &27 & GeAnayet & 288&g & 30&83 \\
FFC/FCP connector, Hirose & 28 & Paquito & 19 pc&(1.23 g/pc)&  6&83\\
P5K connector, Panasonic  & 29 & Paquito & 15 pc&(0.67 g/pc) & 7&58 \\
Thermoplastic connector, Molex & 30 & GeLatuca & 29 pc&(0.53 g/pc) & 17&20 \\
Solder paste, Multicore & 31& GeLatuca & 457&g & 44&30 \\
Solder wire, Multicore & 32& Paquito & 91&g & 7&74 \\
Ta capacitor, Vishay Sprague  & 33 & GeAnayet & 277 pc&(0.64 g/pc) & 17&97 \\
 \hline
%\normalsize
\end{tabular}
 \label{gedata}
%\end{center}
\end{table}

\footnotesize
\begin{landscape}
%\begin{center}
 \label{rpm}
\begin{longtable}{p{0.2cm}p{3.2cm}p{3.1cm}p{1.2cm}p{1cm}p{1cm}p{1.3cm}p{1.3cm}p{1.3cm}p{1.3cm}p{1.2cm}p{1cm}p{1cm}}

\hline
\textbf{\#} & \textbf{Material} & \textbf{Supplier} & \textbf{Technique} & \textbf{Unit} & \textbf{$^{238}$U} & \textbf{$^{226}$Ra} & \textbf{$^{232}$Th} &\textbf{$^{228}$Th} & \textbf{$^{235}$U}& \textbf{$^{40}$K}  & \textbf{$^{60}$Co}& \textbf{$^{137}$Cs}\\
 \hline
%\hline
\endfirsthead
(Continuation)\\
\hline
\textbf{\#} & \textbf{Material} & \textbf{Supplier} & \textbf{Technique} & \textbf{Unit} & \textbf{$^{238}$U} & \textbf{$^{226}$Ra} &\textbf{$^{232}$Th} &\textbf{$^{228}$Th} & \textbf{$^{235}$U}& \textbf{$^{40}$K}  & \textbf{$^{60}$Co}& \textbf{$^{137}$Cs}\\
\hline
%\hline
\endhead
&(Follows at next page)\\
\endfoot

%&(a) Average on different isotopes\\
%&(b) Activity of $^{226}$Ra\\
%&(c) Two values for early/late part of chain\\
%&(d) Activity of $^{228}$Th\\
%&(e) Activity from $^{214}$Bi\\
%&(f) Activity from $^{208}$Tl\\
%&(g) Special fabrication\\
%&(*) Level obtained from the minimum detectable activity of the detector (MDA)\\
%&(+) Preliminary result.\\
\endlastfoot

 &\textbf{Shielding}&&&&&&&&&&&\\ \hline
1 & Pb & Cometa& GDMS & mBq/kg & 0.37&& 0.073 &&& $<$0.31 && \\
2 & Pb & Mifer& GDMS & mBq/kg&$<$1.2&& $<$0.41 &&&0.31  && \\
3 & Pb &Mifer & GDMS & mBq/kg&0.33&& 0.10 &&& 1.2 && \\ %(new supplier)
4 & Pb &Tecnibusa & GDMS & mBq/kg&0.73&& 0.14 &&& 0.91 && \\
5 & Pb &Tecnibusa & Ge & mBq/kg & $<$94 & $<$2.0& $<$3.8 & $<$4.4 &
$<$30 & $<$2.8 & $<$0.2 & $<$0.8\\
6 & Pb &Tecnibusa & Ge & mBq/kg & $<$57 & $<$1.9& $<$1.7 & $<$2.8 &
$<$22 & $<$1.7 & $<$0.1 & $<$0.5\\
%\hline

7  & Cu (ETP) & Sanmetal & GDMS & mBq/kg & $<$0.062 && $<$0.020 &&&     && \\
8  & Cu (C10100) & Luvata (hot rolled) & GDMS &mBq/kg& $<$0.012&& $<$0.0041 &&& 0.061 && \\
9  & Cu (C10100) & Luvata (cold rolled) & GDMS &mBq/kg& $<$0.012&& $<$0.0041 &&& 0.091 && \\
10  & Cu (C10100) & Luvata (hot+cold rolled)& Ge & mBq/kg& &$<$7.4 &
 $<$0.8 & $<$4.3 && $<$18 & $<$0.8 & $<$1.2 \\
\hline

 &\textbf{Vessel}&&&&&&&&&&&\\ \hline
11 & Ti &SMP & Ge  & mBq/kg & $<$233 & $<$5.7&$<$8.8&$<$9.5&3.4$\pm$1.0& $<$22 & $<$3.3 & $<$5.2\\
12 &  Ti &SMP & Ge  & mBq/kg & $<$361 & $<$6.6& $<$11 & $<$10 & $<$8.0& $<$15 & $<$1.0 & $<$1.8 \\
13 & Ti &Ti Metal Supply & Ge  & mBq/kg & $<$14 & $<$0.22& $<$0.5 & 3.6$\pm$0.2 & 0.43$\pm$0.08& $<$0.6 & $<$0.07 & $<$0.07\\
 %\hline

14 & 304L SS& Pfeiffer  & Ge & mBq/kg& & 14.3$\pm$2.8 & 9.7$\pm$2.3& 16.2$\pm$3.9& 3.2$\pm$1.1 & $<$17 & 11.3$\pm$2.7 & $<$1.6\\
15 & 316Ti SS & Nironit, 10-mm-thick & Ge  & mBq/kg & $<$21 & $<$0.57 & $<$0.59 & $<$0.54 & $<$0.74  & $<$0.96 & 2.8$\pm$0.2  & $<$0.12  \\
16 & 316Ti SS & Nironit, 15-mm-thick & Ge  & mBq/kg & $<$25 & $<$0.46 & $<$0.69 & $<$0.88 & $<$0.75  & $<$1.0 & 4.4$\pm$0.3  & $<$0.17  \\
17 & 316Ti SS & Nironit, 50-mm-thick & Ge  & mBq/kg & 67$\pm$22 & $<$1.7 & 2.1$\pm$0.4 & 2.0$\pm$0.7 & 2.4$\pm$0.6  & $<$2.5 & 4.2$\pm$0.3  & $<$0.6  \\
%\hline

18& Inconel 625 & Mecanizados Kanter & Ge & mBq/kg & $<$120 & $<$1.9
& $<$3.4 &
$<$3.2 & $<$4.6 &$<$3.9 & $<$0.4 & $<$0.6\\
19& Inconel 718 & Mecanizados Kanter & Ge & mBq/kg & 309$\pm$78 &
$<$3.4 & $<$5.1 & $<$4.4 & 15.0$\pm$1.9 &$<$13 & $<$1.4 & $<$1.3 \\
\hline

 &\textbf{HV, EL components}&&&&&&&&&&&\\ \hline
20 & PEEK & Sanmetal & Ge  & mBq/kg& & 36.3$\pm$4.3 & 14.9$\pm$5.3&
11.0$\pm$2.4 & $<$7.8 & 8.3$\pm$3.0 & $<$3.3 & $<$2.6 \\
21 & Polyethylene & IN2 Plastics & Ge & mBq/kg & $<$140 & $<$1.9 &
$<$3.8 & $<$2.7 & $<$1.0 & $<$8.9 & $<$0.5 & $<$0.5 \\
22 & Semitron ES225 & Quadrant EPP & Ge & mBq/kg & $<$101 & $<$2.3 &
$<$2.0 &
$<$1.8 & 1.8$\pm$0.3 & 513$\pm$52 & $<$0.5 & $<$0.6 \\
23 & SMD resistor & Farnell & Ge  & mBq/pc& 2.3$\pm$1.0 &
0.16$\pm$0.03 & 0.30$\pm$0.06 & 0.30$\pm$0.05  & $<$0.05 & 0.19$\pm$0.08 & $<$0.02 & $<$0.03 \\
24 & SM5D resistor & Finechem & Ge  & mBq/pc& 0.4$\pm$0.2 &
0.022$\pm$0.007 & $<$0.023 & $<$0.016  & 0.012$\pm$0.005 & 0.17$\pm$0.07& $<$0.005 & $<$0.005 \\
\hline

&\textbf{Energy, tracking planes}&&&&&&&&&&&\\
\hline

25 & Kapton-Cu PCB & LabCircuits & Ge & mBq/cm$^{2}$ & $<$0.26 & $<$0.014 & $<$0.012 & $<$0.008  & $<$0.002 & $<$0.040 & $<$0.002 & $<$0.002  \\
26 & Cuflon & Polyflon & Ge & mBq/kg & $<$33 & $<$1.3 & $<$1.1 & $<$1.1  & $<$0.6 & 4.8$\pm$1.1 & $<$0.3 & $<$0.3  \\
27 & Bonding films & Polyflon & Ge & mBq/kg & 1140$\pm$300 &
487$\pm$23 & 79.8$\pm$6.6 &
66.0$\pm$4.8 & 60.0$\pm$5.5 & 832 $\pm$87 & $<$4.4 & $<$3.8 \\
28& FFC/FCP connector & Hirose & Ge & mBq/pc & $<$50 & 4.6$\pm$0.7 & 6.5$\pm$1.2 &6.4$\pm$1.0 & $<$0.75 & 3.9$\pm$1.4 & $<$0.2 & $<$0.5  \\
29& P5K connector & Panasonic & Ge& mBq/pc & $<$42
& 6.0$\pm$0.9 & 9.5$\pm$1.7 & 9.4$\pm$1.4 & $<$0.95 & 4.1$\pm$1.5 & $<$0.2 & $<$0.8\\
30 & Thermopl. connector & Molex & Ge & mBq/pc & $<$7.3
& 1.77$\pm$0.08 & 3.01$\pm$0.19 & 2.82$\pm$0.15 & $<$0.31 & 2.12$\pm$0.25 & $<$0.022 & 0.27$\pm$0.03\\
31 & Solder paste & Multicore & Ge & mBq/kg & $<$310 & $<$4.9 &
$<$8.0
& $<$6.0 & $<$5.2 & $<$13 & $<$1.0 & $<$1.6 \\
32& Solder wire & Multicore & Ge & mBq/kg & $<$4900 &
(7.7$\pm$1.2)10$^{2}$ &
$<$147 & $<$14 & & $<$257 & $<$30 & $<$36 \\
33 & Ta capacitor & Vishay Sprague & Ge & mBq/pc & $<$0.8 &
0.043$\pm$0.003 & 0.034$\pm$0.004 & 0.032$\pm$0.003 & $<$ 0.010 & & $<$0.002  & $<$0.003 \\
\hline

\caption{Activities measured in relevant materials for NEXT and
following different techniques. GDMS results were derived from U, Th
and K concentrations. Germanium gamma-ray spectrometry results
reported for $^{238}$U and $^{232}$Th correspond to the upper part
of the chains and those of $^{226}$Ra and $^{228}$Th give activities
of the lower parts (see text).}

\end{longtable}
%\label{rpm}
%\end{center}
\end{landscape}
\normalsize

%cleaning protocol email Sara 1/6/12
For the metal samples prepared for germanium spectrometry, a
standard cleaning protocol was followed including the following
steps: diamond cut, cleaning with acetone (alcohol was avoided
because it may affect their mechanical properties), ultrasound bath
with acid-detergent, 63\% nitric acid bath, new cleaning with
acetone and storing in a sealed plastic bag. Other samples were
cleaned in an ultrasonic bath and with pure alcohol.

In the following, results are presented and discussed for the
different screened materials grouped according to their position in
the detector set-up.

\subsection{Shielding}

The external passive shielding for NEXT-100 will be made of lead
bricks forming a 20-cm-thick lead castle and there will be also an
additional 12-cm-thick inner layer of copper to attenuate the
radiation originated in the vessel material \cite{jinsttdr}. Copper
is expected to shield in-vessel electronics components if necessary
and for the photomultipliers enclosures too.

%lead
Lead samples from different suppliers
(Mifer\footnote{http://www.mifer.com}, using two different raw
materials, and Tecnibusa\footnote{http://www.tecnibusa.com} from
Spain and COMETA\footnote{http://www fonderiaroma.com} from Italy)
were screened by GDMS. Results are shown in rows \#1-4 of table
\ref{rpm}; it must be noted that the quantified U and Th
concentrations were reported to be at the ultimate limit of
detection. The results obtained for COMETA lead are in agreement
with the specifications given by the company.
% 2 & Pb, Cometa & & mBq/kg & $<$0.5 & $<$0.5 & $<$1 & Cometa \\
%and offer $<$100 mBq/kg of $^{210}$Pb.
Since a large amount of the lead bricks will be ultimately provided
by Tecnibusa, two different half-brick samples
(10$\times$10$\times$5 cm$^{3}$ each) from this company were
measured at LSC; results are presented in rows \#5-6 of table
\ref{rpm}. This lead has a low activity of $^{210}$Pb at the level
of some tens of Bq/kg. U and Th contamination in lead are normally
not very important \cite{heusser}, since radioactive contaminants
are effectively removed from lead together with silver \cite{ale91}.
For instance, for Dow Run lead produced by JL
Goslar\footnote{http://www.doerun.com, http://www.jlgoslar.de},
activities of tens of $\mu$Bq/kg were measured in \cite{ARI04} and
even lower values have been presented by EXO \cite{LEO08} and GERDA
\cite{BUD08} experiments as upper limits.
%resultados de Pb210 sin incluir

%copper
Three copper samples having different origins were also screened by
GDMS. One is Electrolytic Tough Pitch (ETP) copper supplied by the
Spanish company Sanmetal\footnote{http://www.sanmetal.es} while the
other two were made of C10100 copper from the Luvata
company\footnote{http://www.luvata.com}, having different production
mechanism (hot versus cold rolling). The Luvata copper samples were
screened together using the Paquito detector as well. All results on
copper are shown in rows \#7-10 of table \ref{rpm}. The upper bounds
on activities derived from the germanium spectrometry measurement
were much less stringent than those from GDMS due to its limited
sensitivity; hence a new measurement, with much more mass and time
and using a bigger germanium detector is foreseen. The cleanest
copper we are aware of is that supplied by Norddeutsche Affinerie
(Germany)\footnote{Now re-branded as Aurubis,
http://www.aurubis.com}; very low upper limits for its activity were
set in measurements at the Gran Sasso Underground Laboratory
\cite{ARI04} and by the EXO Collaboration \cite{LEO08} and even
activity from the natural chains and $^{40}$K was quantified by the
XENON experiment \cite{APR11}, at levels of a few tens of
$\mu$Bq/kg. Although the GDMS measurement of Luvata copper has given
information only on U and Th concentration, the upper limits derived
are at the same level or even better than the results for the
Norddeutsche Affinerie copper, and Luvata copper has therefore been
chosen as the first option for the NEXT shield.

\subsection{Vessel}

The pressure vessel of NEXT-100 must be able to hold 15 bar of
xenon. It consists of a cylindrical center section (barrel) with two
identical torispherical heads on each end \cite{jinsttdr}. The
vessel orientation is horizontal, so as to minimize the overall
height. Although it will be ultimately made of stainless steel, the
first considered option was titanium, so several samples of both
materials have been screened. Inconel (nickel-chromium alloy) will
be used to bolt the end-caps to the main body due to its excellent
strength properties and therefore its radiopurity has been analyzed
too.

%% Titanium
Grade 2 titanium was initially proposed to be used for the main
components of the vessel. Two samples were screened at LSC, one from
a Spanish supplier, Titanio SMP\footnote{http://www.titaniosmp.com}
and the other from Titanium Metal
Supply\footnote{http://www.titaniummetalsupply.com}. Results are
shown in rows \#11-13 of table \ref{rpm}. The Ti SMP sample was
screened using two different germanium detectors, GeOroel (row \#11)
and GeTobazo (row \#12) as a cross-check exercise; the small
differences found can be well understood taking into account the
differences in the background rates of the detectors in several
energy ranges. Thanks to the much larger mass available in the
sample, upper limits on activities derived for the Ti Metal Supply
sample are much lower than for the Ti SMP sample and it has been
possible to quantify the activity of the lower part of the
$^{232}$Th chain. Production of $^{46}$Sc, beta emitter with
Q$=$2366.7 keV and T$_{1/2}=$83.8 days, has been also observed for
this sample; it must have been generated by (n,p) reactions on
$^{46}$Ti induced by fast neutrons. The LUX collaboration has
carried out an exhaustive analysis of Ti samples \cite{LUX11},
obtaining different levels of radiopurity for them; the presence of
$^{46}$Sc is usually identified.

%% Stainless steel
A great deal of activity measurements for different types of
stainless steel (SS) can be found in the literature (see for example
Refs. \cite{BOR,SNO,UKDMC,ILI11}) showing a wide range of values.
One sample of type 304L (a vacuum system piece from
Pfeiffer\footnote{http://www.pfeiffer-vacuum.com}) was screened
using the Paquito detector (see results in row \#14 of table
\ref{rpm}), obtaining the usual quite high levels of activity from
natural chains. Material referenced as austenitic 1.4571 (also 316L)
has been extensively studied by XENON \cite{APR11} and GERDA
\cite{MAN08} experiments, finding materials supplied by the Nironit
company\footnote{http://www.nironit.de} with activity levels (values
or upper limits) of even tenths of mBq/kg for isotopes from the
natural chains in the best cases. Three samples of 316Ti stainless
steel supplied by Nironit were screened at LSC and results are
presented at rows \#15-17 of table \ref{rpm}; they have different
thickness since they are intended to be used in different parts of
the NEXT pressure vessel (10 mm for body, 15 mm for end-caps and 50
mm for flanges). Activities from $^{60}$Co and $^{54}$Mn, commonly
present in steel, have been quantified for the three samples.
Results for cosmogenic $^{54}$Mn, not quoted in table \ref{rpm}, are
0.29$\pm$0.05, 0.55$\pm$0.07 and 0.97$\pm$0.14 mBq/kg for increasing
thickness of sample. For the 10- and 15-mm-thick samples, upper
bounds for all the other emitters investigated have been derived;
for the 50-mm-thick sample, the activity of the isotopes of the
$^{232}$Th chain has been quantified, pointing to secular
equilibrium. It is worth noting that the sensitivity for the
thickest sample was worse than for the other two, because of the
lower mass available in the sample (see table \ref{gedata}) and the
lower efficiency detection due to self-absorption. The activity
values obtained for Nironit 316Ti stainless steel are of the order
or below NEXT requirements; therefore, the booked batches from where
the samples were taken will be used for pressure vessel
construction.
%comparacion GERDA; XENON

%Electron beam, TIG or MIG are possibilitis for the welding. A
%sample referenced as ``Stainless steel TIG'' from Harris Product
%Group has been screened at SNOlab and results are quoted in row \#26
%in Table \ref{rpm}. EXO has reported (measurement \#6 at
%\cite{LEO08}) the concentrations of Th and U after TIG welding on a
%copper sample: $<$9.8 pg/cm of Th, 10.2$\pm$3.4 pg/cm of U. A filler
%for TIG soldering, made of Ti grade 2 (diameter: 2 mm) supplied by
%ETM company is being screened at Canfranc Underground Laboratory. %%% entrada sin incluir en tabla

%%Inconel
Samples of inconel 718 and inconel 625 from the Spanish company
Mecanizados Kanter\footnote{http://www.mecanizados-kanter.es} were
screened at LSC and results are shown in rows \#18-19 of table
\ref{rpm}. No previous results on radiopurity of this material have
been found; upper limits on activities of the lower parts of the
$^{238}$U and $^{232}$Th chains have been set at some mBq/kg for
both types of inconel. Presence  of
$^{58}$Co was identified for inconel 625. %0.8$\pm$0.2

\subsection{High Voltage and Electroluminescence components}

The main body of the {\it field cage} to be placed inside the vessel
will be made of high density polyethylene, with attached copper
strips connected to resistors \cite{jinsttdr}; PEEK was also
considered as an alternative. Wire meshes separating the different
field regions of the detector, including the electroluminescence
volume, will be made of stainless steel. To improve the light
collection efficiency of the detector, reflector panels coated with
a wavelength shifter will cover the inner part of the field cage.
This {\it light tube} will be made of Tetratex$\circledR$ fixed over
a 3M substrate, coated with tetraphenyl butadiene (TPB)
\cite{jinsttdr}. The ArDM experiment has screened specifically
polytetrafluoroethylene (PTFE) Tetratex from Donaldson
Membranes\footnote{http://www.donaldson.com} by Inductively Coupled
Plasma Mass Spectrometry (ICPMS) \cite{BOC09} and TPB from two
different manufacturers were measured at SNOlab \cite{SNO}.

%% peek
A sample of PEEK from Sanmetal Spanish company was screened using
the Paquito detector; values obtained are shown in row \#20 of table
\ref{rpm}, pointing to a non-negligible activity. Only upper bounds
on activity of PEEK were presented in \cite{kimballton}.

%% HDP, plastics
Polyethylene from IN2 Plastics
company\footnote{http://www.in2plastics.com} has a very good
radiopurity according to XENON results \cite{APR11} and a sample of
High Molecular Weight polyethylene (type PE500) was therefore chosen
for screening at LSC. First results, shown in row \#21 of table
\ref{rpm}, have produced only upper limits for common radioisotopes.
Semitron$\circledR$ ES225 plastic produced by Quadrant Engineering
Plastic Products\footnote{http://www.quadrantplastics.com} has been
also measured and the results on its radiopurity are presented in
row \#22 of table \ref{rpm}; in this case, a quite high activity of
$^{40}$K has been registered.

%% resistors
Surface Mount Device (SMD) resistors supplied by
Farnell\footnote{http://www.farnell.com} and by
Finechem\footnote{http://www.jfine.co.jp} were screened at LSC (see
results in rows \#23-24 of table \ref{rpm}). Activity values
obtained for Finechem resistors are much lower than for Farnell
ones, and very similar to those obtained in \cite{APR11} for
resistors of the same type and company.

%\noindent {\bf Polyethylene}: samples of material to be used as
%neutron moderator in shieldings have been screened for different
%dark matter experiments; best results from EDELWEISS (supplier
%Plastiques du Rhone) presented at \cite{ILI11} and from XENON
%(supplier in2plastic, measurement \#14 in Table 1 at \cite{APR11})
%are shown in rows \#36-37 in Table \ref{rpm}. More results are
%available at \cite{ILI11}. Polyethylene insulator from cables has
%been analyzed by EXO and values from measurement \#199 at
%\cite{LEO08} are quoted in row \#38 in Table \ref{rpm}.

\subsection{Energy and tracking planes components}

%% from TDR
The energy measurement in NEXT is provided by the detection of the
electroluminescence light by an array of photomultipliers (the
so-called {\it energy plane}), located behind the cathode mesh.
Those photomultipliers will also record the primary scintillation
light that indicates the start of the event. Each photomultiplier
will be sealed into individual, pressure resistant, vacuum tight
copper enclosures and will be coupled to the sensitive volume
through a sapphire window coated with TPB. The tracking function
will be provided by a plane of MPPCs operating as sensor pixels and
located behind the transparent electroluminescence gap. The MPPCs
will be mounted on square boards named Dice Boards (DB); each DB
will contain 8$\times$8 sensors with a pitch of $\sim$1 cm and will
be coated with TPB too. The array of DBs, with $\sim$7000 channels,
forms the {\it tracking plane}. Front-end electronics boards for
this tracking plane will be inside the pressure vessel, behind a
copper shield. Details can be found in \cite{jinsttdr}.

%Energy plane
%% PMTs, sapphire windows
Although different models of photomultipliers were under
consideration, Hamamatsu R11410MOD was finally chosen for the
NEXT-100 detector since radiopurity specifications provided by
Hamamatsu are better than for other models ($\sim$3.3 mBq/PMT of
$^{238}$U, $\sim$2.3 mBq/PMT of $^{232}$Th and $\sim$5.7 mBq/PMT of
$^{40}$K). The XENON \cite{APR11} and LUX \cite{FAH11} experiments
have studied the radiopurity of this model, finding even lower
activities. These levels would satisfy NEXT requirements, but the
dispersion found in different measurements imposes a dedicated
program to control the photomultiplier radiopurity, which will be
undertaken in LSC. Sapphire windows (from Swiss Jewel
Company\footnote{http://www.swissjewel.com}) were screened by EXO
\cite{LEO08} with activities of tenths of mBq/kg from the natural
chains; a measurement of sapphire windows for NEXT is underway.

% A photomultiplier {\it
%feedthrough} made of steel and glass ceramic, having 41 pins Au
%plated supplied by Ceramtec company has been screened at Canfranc
%Underground Laboratory using GeTobazo detector. Results are in progress. %row \# in Table \ref{rpm}. %%% entrada sin incluir en tabla

%% Tracking plane
%% Si
Silicon, used for the MPPCs at the tracking plane, is, as germanium,
a very radiopure material with typical intrinsic activities of
$^{238}$U and $^{232}$Th at the level of few $\mu$Bq/kg
\cite{heusser}. In any case, a specific measurement of MPPCs is
pending, either separately or after being mounted on the boards.

%% FR4, kapton, cuflon
Several materials have been taken into consideration for Printed
Circuit Boards (PCBs) and a large number of radiopurity measurements
can be found in \cite{ILI11}. FR4 was disregarded for both an
unacceptable high rate of outgassing and bad radiopurity (a fast
measurement made with Paquito detector pointed to activities of
hundreds of mBq/kg for natural chains); it seems that the glass
fiber-reinforced materials at base plates of circuit boards can be a
source of radioactive contamination \cite{heusser}. Components made
of just kapton (like cirlex) and copper offer very good radiopurity,
as shown in the measurements of kapton-copper foils in
\cite{radiopuritymm} and in the screening of a monolayer PCB made of
kapton and copper supplied by
LabCircuits\footnote{http://www.lab-circuits.com} using the Paquito
detector, presented in row \#25 of table \ref{rpm}. Also
cuflon$\circledR$ offers low activity levels, as shown in the
measurement of samples from Crane
Polyflon\footnote{http://www.polyflon.com} by GERDA \cite{BUD09} and
at \cite{NIS09}, using both ICPMS and Ge gamma spectroscopy. A first
measurement of Polyflon cuflon made of a 3.18-mm-thick PTFE layer
sandwiched by two 35-$\mu$m-thick copper sheets was made for NEXT
and results are shown in row \#26 of table \ref{rpm}. Cuflon has
been chosen for the Dice Boards. Adhesive films to glue cuflon
sheets are used to prepare multilayer PCBs; a sample of bonding
films made of a polyolefin co-polymer and supplied also by Crane
Polyflon were screened and results are presented in row \#27 of
table \ref{rpm}. Despite the high activities measured, the use of
these films is not completely discarded since the necessary mass for
each Dice Board is of only 0.59 g.

%% Connectors, solder
Information is also available on the radiopurity of different types
of connectors \cite{BOR,SNO,ILI11}. FFC/FCP (Flexible Printed
Circuit \& Flexible Flat Cable) connectors were initially considered
for the DBs; connectors supplied by
Hirose\footnote{http://www.hirose.com} were selected and screened
using the Paquito detector, as well as similar connectors from
Panasonic\footnote{http://www.panasonic-electric-works.com} (see
results in rows \#28-29 of table \ref{rpm}). Both types of
connectors show activities of at least several mBq/pc for isotopes
in $^{232}$Th and the lower part of $^{238}$U chains and for
$^{40}$K. These values, according to \cite{ILI11}, are of the same
order than those for similar connectors supplied by
Molex\footnote{http://www.molex.com}; all of them are made of Liquid
Crystal Polymer (LCP), thus it seems that the activity measured is
related to this material. Thermoplastic connectors 503066-8011 from
Molex were also screened (see results in row \#30 of table
\ref{rpm}) finding activities slightly smaller but of the same
order. Since these levels are too high for NEXT sensitivity, it was
concluded that this kind of connectors must be avoided at DBs and
the collaboration is at the moment foreseeing a direct bonding of
the FFCs/FCPs to the DBs. A sample of lead free SnAgCu solder paste
supplied by Multicore (Ref. 698840) has been screened and results
are presented in row \#31 of table \ref{rpm}). $^{108m}$Ag, induced
by neutron interactions and having a half-life of T$_{1/2}=$438 y,
has been identified in the paste, with an activity of 5.26$\pm$0.40
mBq/kg, while upper limits of a few mBq/kg have been set for the
common radioactive isotopes; consequently, this solder paste could
be used without concern. Solder wire with similar composition from
Multicore (Ref. 442578) was also screened (see row \#32 of table
\ref{rpm}), finding in this case a high activity of the lower part
of the $^{238}$U chain. An activity of $^{210}$Pb of
(1.2$\pm$0.4)$\times$10$^{3}$ Bq/kg has been deduced using the
bremsstrahlung emission from its daughter nuclide $^{210}$Bi
\cite{nachab}.

%% Capacitors
Concerning electronic components at the boards, ceramic capacitors
seem to be quite radioactive \cite{ILI11} and have been disregarded.
Tantalum capacitors (Vishay Sprague
597D\footnote{http://www.vishay.com}) have been screened at LSC and
results are presented in row \#33 of table \ref{rpm}; activity
levels are lower than for tantalum capacitors supplied by
AVX\footnote{http://www.avx.com}, from database in \cite{ILI11}. In
addition to activities shown in table \ref{rpm},
%a level of 0.0010$\pm$0.0003 mBq/pc of $^{54}$Mn has been measured and
the presence of $^{182}$Ta (beta emitter with Q$=$1814.3 keV and
T$_{1/2}=($114.74$\pm$0.12) days \cite{singh}, produced by neutron
activation on $^{181}$Ta) has been identified through the
observation of many of its gamma emissions.

\section{Summary}

A thorough radiopurity control is being performed for the
construction of the NEXT-100 double beta decay experiment to be
operated at LSC. Evaluation of activity levels in the most relevant
materials has been carried out based on GDMS and on ultra-low
background germanium gamma-ray spectrometry at LSC. Some of the
results presented here represents the first measurements performed
using the germanium detectors of the LSC Radiopurity Service. This
kind of data have a widespread interest, being also useful for other
experiments investigating rare phenomena in underground
laboratories. Radiopurity information is expected to be helpful not
only for the selection of materials that are radiopure enough to
minimize the background level, but also for the development of the
NEXT detector background model in combination with Monte Carlo
simulations.

Adequate materials for the external and internal passive shields
have been identified. Titanium and stainless steel samples were
analyzed for the construction of the pressure vessel. The good
radiopurity found for the 316Ti stainless steel supplied by the
Nironit company, of the order of tenths of mBq/kg for the isotopes
at the lower part of the $^{238}$U and $^{232}$Th chains, confirmed
that this material can be used for the detector vessel in
combination with an inner copper shield. The background contribution
of the vessel and both internal and external shields has been
studied, assuming the upper limits on the activities obtained here
for Nironit 316Ti stainless steel, Luvata copper and Tecnibusa lead;
the preliminary results indicate, for each material, a background
level in the RoI at most of the order of $10^{-4}$ counts keV$^{-1}$
kg$^{-1}$ y$^{-1}$, after all cuts (fiducial and topological).
Construction of shielding and vessel is thus proceeding.

Selection of in-vessel components to be used at the energy and
tracking planes has been performed too, helping in the design of DB
and front-end electronics boards. While resistors, capacitors and
solder paste of acceptable radiopurity have been found,
board-to-cable connectors containing LCP have been discarded since
they have activities of few mBq per piece for isotopes of the
$^{238}$U and $^{232}$Th chains. Cuflon, being radiopure enough, has
been chosen for the construction of the boards for the MPPCs.

Further radiopurity measurements using germanium detectors at LSC
are scheduled before the installation and operation of the NEXT
detector with the aim to control all the components that might
contribute significantly to the background. Main efforts will be
devoted to the careful selection of photomultipliers of the chosen
model (Hamamatsu R11410MOD) and to the screening of related
components (e.g., saphire windows, bases and cans) at the energy
plane and light tube materials. This task is essential to achieve
the required background level in the region of interest imposed by
the NEXT sensitivity goal.

 \acknowledgments We deeply acknowledge LSC
directorate and staff for their strong support for performing the
measurements at the LSC Radiopurity Service. The NEXT Collaboration
acknowledges funding support from the following agencies and
institutions: the Spanish Ministerio de Economía y Competitividad
under grants CONSOLIDER-Ingenio 2010 CSD2008-0037 (CUP),
Consolider-Ingenio 2010 CSD2007- 00042 (CPAN), and under contracts
ref. FPA2008-03456, FPA2009-13697-C04-04; FCT(Lisbon) and FEDER
under grant PTDC/FIS/103860/2008; the European Commission under the
European Research Council T-REX Starting Grant ref.
ERC-2009-StG-240054 of the IDEAS program of the 7th EU Framework
Program; Director, Office of Science, Office of Basic Energy
Sciences, of the US Department of Energy under contract no.
DE-AC02-05CH11231. Part of these grants are funded by the European
Regional Development Fund (ERDF/FEDER). J. Renner (LBNL)
acknowledges the support of a US DOE NNSA Stewardship Science
Graduate Fellowship under contract no. DE-FC52-08NA28752.
F. I. acknowledges the support from the Eurotalents program. %[to be completed]

\end{document}